# Positive and negative word of mouth in the United States

Shawn Berry, DBA[1]*

[1]William Howard Taft University, Lakewood CO, USA

*Correspondence: <u>shawn.berry.8826@taftu.edu</u>

**Abstract** Word of mouth is a process by which consumers transmit positive or negative sentiment to other consumers about a business. While this process has long been recognized as a type of promotion for businesses, the value of word of mouth is questionable. This study will examine the various correlates of word of mouth to demographic variables, including the role of the trust of business owners. Education level, region of residence, and income level were found to be significant predictors of positive word of mouth. Although the results generally suggest that the majority of respondents do not engage in word of mouth, there are valuable insights to be learned.

---

## 1. Introduction

Word of mouth (WOM) has long been recognized as an important factor to influence consumer behavior. Bonfrer (2011) states that "word of mouth is defined as some type of communication between two or more connected customers or potential customers. Such communication can take the form of verbal, face-to-face communication or many other means, including e-mail, telephone, cell-phone-based text messages, blogs, instant messaging, social network software (e.g., Facebook, Friendster, LinkedIn), and so on" (p.325). Toyama (2020) frames positive word of mouth in terms of consumer satisfaction, stating "consumers' intentions toward positive word-of-mouth are thought to increase when they expect their motivations to be satisfied" (p.25), and noting the role of consumer delight in positive word of mouth, which is "the surprising, positive emotional response that occurs when a consumer's experience far exceeds expectations" (Toyama, 2020, p.25). Dalzotto et al. (2016) define "negative word-of-mouth (WOM)... as derogatory information disseminated from person to person and aimed at defaming a product, highlighting a product complaint, and/or highlighting unsatisfactory service experiences" (p.418). The extent to which consumers engage in word of mouth is varied. observe that "word-of-mouth interactions may involve underreporting (not everyone shares experiences), positivity (positive experiences are communicated more widely than negative ones), or negativity (negative experiences are communicated more widely than positive ones)" (p.168). Naylor and Kleiser (2000) found that "more individuals engaged in positive than negative WOM" (p.26).

Word of mouth has serious implications for businesses. First, word of mouth can persuade consumers. Charlett, Garland, and Marr (1995) observed "that WOM, both positive and negative, is indeed a force that can influence the attitudes and predicted purchase behavior of consumers" (p.42). Halstead (2002) found "that negative word of mouth was greatest among consumers who had also voiced complaints to the seller" (p.1). Second, the reputation of a company or brand can be helped or harmed by word of mouth, depending on whether it is positive or negative (Dalzotto et al. 2016). Third, the loyalty of customers is affected. Manyanga, Makanyeza, and Muranda (2022) observed that "customer experience, satisfaction and word-of-mouth intention were found to have a direct positive effect on loyalty" (p.1). Marcos and Coelho (2022) remark that "the relationship between loyalty and WOM has been poorly studied, although it is well known that in services, loyal customers speak well of the companies" (p.957).

Scholars have generally found relationships between demographic variables and word of mouth, both positive and negative. Falls and Francis (1996) observed that "sex, educational level, and age were found to be effective in predicting positive WOM (complimenting), and household income and age were found to be effective in predicting negative WOM (complaining)" (p.206). Singh and Sarma (2018) found "that word of mouth outcomes differ across different genders, age groups, educational levels, occupation categories, income groups and family sizes" (p.71). In Czechia, Mladenović, Bruni, and Kalia (2021) state that "our findings indicate that offline context is preferred, trust and social ties are positively related to WOM, whilst homophily has reportedly negative impact. Age, gender, and education are reported to influence consumers' propensity to engage in WOM" (p.418).

The propensity to engage in positive and negative word of mouth will be explored in this study. Using data from my dissertation (Berry, 2024a), word of mouth behavior will be analyzed according to the demographic characteristics of respondents with respect to age, gender, level of education, level of income, and region of the United States. Additionally, word of mouth behavior will be analyzed according to the trust of business owners by respondents as a general proxy for consumer loyalty, as measured with the trust of people instrument in Berry (2024a). These valuable insights will add to the body of knowledge with respect to word of mouth behavior, and assist those in the field of marketing research. The study will conclude with recommendations and directions for future research.

## 2. Materials and Methods

The data for this study was collected from male and female users of Amazon Mechanical Turk (mTurk) residing in the United States by way of an online questionnaire that was used for my dissertation (Berry, 2024a). The questionnaire was designed to collect a variety of information from respondents that related to their demographic characteristics, habits, beliefs, and attitudes concerning their consumption and posting of online reviews (Berry, 2024b). Of the 398 respondents, "11.8% of the 398 respondents were excluded due to not consenting to the study (3), self-reporting as not residing in the United States (8), attention check question failure (24), and identification of attempts to take the survey more than once (12)" (p.3), resulting in n=351 n=351 (Berry, 2024b).

The trust of business owners by respondents was evaluated using responses to an instrument that quantifies the trust of people in Berry (2024a) by rating a collection of individuals in varying roles in society with a 5-point Likert scale (Berry, 2024b). The instrument was evaluated with Cronbach's alpha, the value of which "was 0.85 (Berry, 2024), which is considered to be high (Taber, 2018), and therefore, reliable" (Berry, 2024b, p.3).

The data was coded for analysis according to the scheme as illustrated in Table 1 below.

**Table** 1

*Variable coding scheme*

| Variable name | Variable description | Variable type | Coding |
|---|---|---|---|
| Age | Age group of respondent, years of age | Categorical | 18-24 = 1<br>25-34 = 2<br>35-44 = 3<br>45-54 = 4<br>55 and over = 5 |
| Gender | Gender of respondent | Categorical | Female = 0<br>Male = 1<br>Non-binary = 2 |
| Income | Annual income level of respondent, USD | Categorical | Less than $30,000 = 1<br>$30,000-$49,999 = 2<br>$50,000-$69,999 = 3<br>$70,000 and over = 4 |
| Education | Education level of respondent | Categorical | Did not finish high school = 0<br>High school graduate = 1<br>Some college = 2<br>Bachelor's degree = 3<br>Master's degree = 4<br>Post-graduate or higher = 5 |
| Region | United States region of residence of respondent | Categorical | Middle Atlantic = 1<br>New England = 2<br>South Atlantic = 3<br>East South Central = 4<br>West South Central = 5<br>Mountain = 6<br>Pacific = 7 |
| Likert scores | Degrees of agreement or importance of behavioral factors to measure trust of people | Ordinal | Definitely distrust = 1<br>Somewhat distrust = 2<br>Neither trust nor distrust = 3<br>Somewhat trust = 4<br>Definitely trust = 5 |

**Source:** Berry (2024b), Table 1.

## 3. Results

The overall frequencies of respondents that engage in word of mouth are illustrated in Table 2. In general, the majority of respondents do not tell others when they have a positive or negative experience. However, 21.9% of respondents tell others after a negative experience as compared to 14.3% of respondents that tell others after a positive experience.

**Table 2**
*Frequency of respondent positive and negative word of mouth according to nature of their experience*

| Nature of experience | Does not tell others | Tells others |
|---|---|---|
| Positive experience | 301 | 50 |
| Negative experience | 274 | 77 |

**Source:** Data analysis, Berry (2024a).

In Berry (2024a), respondents were asked to provide a rating of trust for individuals in different roles, including business owners, using the trust of people instrument. The analysis is displayed in Table 3. Among those respondents that had positive experiences and somewhat and definitely trust business owners, 17.6% would tell others. Among those respondents that had negative experiences and somewhat and definitely distrust business owners, 12,1% would tell others. Among those respondents that held neutral levels of trust toward business owners, 21.9% would tell others about their negative experience and 16.8% would tell others about their positive experience.

**Table 3**
*Frequency of respondent positive and negative word of mouth according to level of trust of business owners*

| Level of trust of business owners | Positive experience Does not tell others | Positive experience Tells others | Negative experience Does not tell others | Negative experience Tells others |
|---|---|---|---|---|
| 1 Definitely distrust | 9 | 4 | 7 | 6 |
| 2 Somewhat distrust | 71 | 7 | 65 | 13 |
| 3 Neiter trust nor distrust | 143 | 24 | 137 | 30 |
| 4 Somewhat trust | 64 | 12 | 65 | 21 |
| 5 Definitely trust | 14 | 3 | 10 | 7 |

**Source:** Data analysis, Berry (2024a).

The frequency of positive and negative word of mouth by respondents according to region of the United States is illustrated in Table 4. Respondents from the Middle Atlantic region had the highest frequency of word of mouth after a positive experience. Respondents from the South Atlantic region had the highest frequency of word of mouth after a negative experience.

With respect to positive word of mouth, the results of the chi square test of independence showed that there is a significant relationship between telling others about a positive experience and the region of residence of respondents, $X2$ (6, $N$ = 351) = 16.557, $p$ = .011. With respect to negative word of mouth, the results of the chi square test of independence showed that there is no significant relationship between telling others about a negative experience and the region of residence of respondents, $X2$ (6, $N$ = 351) = 4.304, $p$ = .636.

**Table 4**
*Frequency of respondent positive and negative word of mouth by region*

| Region | Positive experience Does not tell others | Positive experience Tells others | Negative experience Does not tell others | Negative experience Tells others |
|---|---|---|---|---|
| Middle Atlantic (NY/NJ/PA) | 48 | 19 | 58 | 9 |
| New England (CT/ME/MA/NH/RI//VT) | 16 | 0 | 13 | 3 |
| South Atlantic (DE/DC/FL/GA/MD/NC/SC/VA/WV) | 79 | 11 | 68 | 22 |
| East South Central (AL/KY/MS/TN) | 51 | 6 | 44 | 13 |
| West South Central (AR/LA/OK/TX) | 46 | 4 | 38 | 12 |
| Mountain (AZ/CO/ID/MT/NV/NM/UT/WY) | 25 | 3 | 22 | 6 |
| Pacific (AK/CA/HI/OR/WA) | 36 | 7 | 31 | 12 |

**Source:** Data analysis, Berry (2024a).

The frequency of positive and negative word of mouth by respondents according to age category is illustrated in Table 5. Respondents aged 25 to 34 years had the highest frequency of word of mouth after a positive experience and negative experience. However, those respondents aged 55 and older were more likely to tell others about a positive experience (23.1%). Respondents aged 18 to 24 and those aged 55 and older were more likely to tell others about a negative experience (30.8% and 30.8%, respectively).

With respect to positive word of mouth, the results of the chi square test of independence showed that there is no significant relationship between telling others about a positive experience and the age category of respondents, $X2$ (4, $N = 351$) = 1.744, $p = .783$. With respect to negative word of mouth, the results of the chi square test of independence showed that there is no significant relationship between telling others about a negative experience and the age category of respondents, $X2$ (4, $N = 351$) = 2.339, $p = .674$.

**Table 5**
*Frequency of respondent positive and negative word of mouth by age category*

| Age category | Positive experience Does not tell others | Positive experience Tells others | Negative experience Does not tell others | Negative experience Tells others |
|---|---|---|---|---|
| 18–24 | 21 | 5 | 18 | 8 |
| 25–34 | 119 | 20 | 108 | 31 |
| 35–44 | 100 | 14 | 92 | 22 |
| 45–54 | 51 | 8 | 47 | 12 |
| 55 and older | 10 | 3 | 9 | 4 |

**Source:** Data analysis, Berry (2024a).

The frequency of positive and negative word of mouth by respondents according to level of income is illustrated in Table 6. Respondents earning less than $30,000 per year had the highest frequency of word of mouth after a positive experience (15.9% tell others) and negative experience (26.2% tell others).

With respect to positive word of mouth, the results of the chi square test of independence showed that there is no significant relationship between telling others about a positive experience and the income category of respondents, $X2$ $(3, N = 351) = 2.611$, $p = .456$. With respect to negative word of mouth, the results of the chi square test of independence showed that there is no significant relationship between telling others about a negative experience and the income category of respondents, $X2$ $(3, N = 351) = 2.134$, $p = .545$.

**Table 6**
*Frequency of respondent positive and negative word of mouth by level of income*

| Income category | Positive experience Does not tell others | Positive experience Tells others | Negative experience Does not tell others | Negative experience Tells others |
|---|---|---|---|---|
| Less than $30,000 | 106 | 20 | 93 | 33 |
| $30,000–$49,999 | 88 | 18 | 86 | 20 |
| $50,000–$69,999 | 44 | 5 | 39 | 10 |
| $70,000 and over | 63 | 7 | 56 | 1 |

**Source:** Data analysis, Berry (2024a)
.

The frequency of positive and negative word of mouth by respondents according to level of education is illustrated in Table 7. Respondents possessing a master's degree had the highest frequency of word of mouth after a positive experience (38.8% tell others). Respondents possessing a bachelor's degree had the highest frequency of word of mouth after a negative experience (28.7% tell others). However, among those respondents that did not finish high school, 100% told others about their negative experience.

With respect to positive word of mouth, the results of the chi square test of independence showed that there is a significant relationship between telling others about a positive experience and the level of education of respondents, $X2$ (5, $N = 351$) = 30.908, $p < 0.001$. With respect to negative word of mouth, the results of the chi square test of independence showed that there is no significant relationship between telling others about a negative experience and the income category of respondents, $X2$ (5, $N = 351$) = 8.643, $p = .124$.

**Table 7**
*Frequency of respondent positive and negative word of mouth by level of education*

| Level of education | Positive experience Does not tell others | Positive experience Tells others | Negative experience Does not tell others | Negative experience Tells others |
|---|---|---|---|---|
| Did not finish high school | 1 | 1 | 0 | 2 |
| High school graduate | 42 | 4 | 36 | 10 |
| Some college | 114 | 14 | 103 | 25 |
| Bachelor's degree | 101 | 11 | 87 | 35 |
| Master's degree | 30 | 19 | 36 | 13 |
| Postgraduate or higher | 13 | 1 | 12 | 2 |

**Source:** Data analysis, Berry (2024a).

The frequency of positive and negative word of mouth by respondents according to their gender is illustrated in Table 8. Although female respondents had the highest frequency of word of mouth after a positive experience, 17.6% of male respondents tell others about their positive experience. Female respondents had the highest frequency of word of mouth after a negative experience (22.1% tell others). Of non-binary respondents that had negative experiences, 50% of them tell others about it.

With respect to positive word of mouth, the results of the chi square test of independence showed that there is no significant relationship between telling others about a positive experience and the gender of respondents, $X2$ $(2, N = 351) = 1.836$, $p = .399$. With respect to negative word of mouth, the results of the chi square test of independence showed that there is no significant relationship between telling others about a negative experience and the gender of respondents, $X2$ $(2, N = 351) = .256$, $p = .879$.

**Table 8**
*Frequency of respondent positive and negative word of mouth by gender*

| Gender | Positive experience Does not tell others | Positive experience Tells others | Negative experience Does not tell others | Negative experience Tells others |
|---|---|---|---|---|
| Female | 209 | 31 | 187 | 53 |
| Male | 89 | 19 | 85 | 23 |
| Non-binary | 3 | 0 | 2 | 1 |

**Source:** Data analysis, Berry (2024a).

The data was analyzed using logistic regression to model positive word of mouth using the demographic variables. The dependent variable *wom-positive* is a binary variable, and equal to one if the respondent tells others about their positive experience or zero if they do not. The model results are displayed in Table 9. The Akaike Information Criterion (AIC) for the model was 284.45. Age and gender were not statistically significant predictors of positive word of mouth. The intercept and level of education of respondents were highly statistically significant, followed by level of income, and to a lesser extent, the region of residence for respondents. Gender, age, and the level of trust of business owners by respondents were not statistically significant. As the education level increases, this strongly increases the likelihood of positive word of mouth. As the level of income strongly decreases the likelihood of positive word of mouth.

**Table 9**

*Logistic regression model of positive word of mouth by respondents (dependent variable: wom-positive)*

|  | Estimate | Std. Error | z value | Pr(|z|) | Significance |
|---|---|---|---|---|---|
| Intercept | -1.770 | 0.794 | -2.229 | 0.026 | (*) |
| Gender | 0.345 | 0.318 | 1.086 | 0.278 |  |
| Income | -0.378 | 0.158 | -2.391 | 0.017 | (*) |
| Region | -0.149 | 0.082 | -1.831 | 0.067 | (#) |
| Education | 0.499 | 0.161 | 3.104 | 0.002 | (**) |
| Age | -0.064 | 0.171 | -0.372 | 0.709 |  |
| Business owner trust | -0.001 | 0.184 | -0.008 | 0.994 |  |

$^{\#}p < 0.1$. $^{*}p < .05$. $^{**}p < .01$. $^{***}p < .001$.

**Source:** Data analysis, Berry (2024a)

Similar to the analysis in Table 9, logistic regression was used to create a model for negative word of mouth using the demographic variables. The dependent variable *wom-negative* is a binary variable, and equal to one if the respondent tells others about their positive experience or zero if they do not. The model results are displayed in Table 10. The Akaike Information Criterion (AIC) for the model was 377.97. Although the intercept was statistically significant, all of the predictor variables were not statistically significant. The trust of business owners and region of residence appear to increase the likelihood of negative word of mouth among respondents.

**Table 10**

*Logistic regression model of negative word of mouth by respondents (dependent variable: wom-negative)*

| | Estimate | Std. Error | z value | Pr(|z|) | Significance |
|---|---|---|---|---|---|
| Intercept | -1.904 | 0.685 | -2.779 | 0.006 | (**) |
| Gender | 0.029 | 0.269 | 0.106 | 0.915 | |
| Income | -0.102 | 0.128 | -0.796 | 0.426 | |
| Region | 0.107 | 0.067 | 1.573 | 0.116 | |
| Education | 0.018 | 0.137 | 0.129 | 0.898 | |
| Age | -0.071 | 0.142 | -0.501 | 0.616 | |
| Business owner trust | 0.188 | 0.148 | 1.269 | 0.205 | |

$^\#p <0.1.$ $^*p < .05.$ $^{**}p < .01.$ $^{***}p < .001.$

**Source:** Data analysis, Berry (2024a)

## 4. Discussion

The findings of this study primarily suggest that a large contingent of consumers don't actively engage in word of mouth to a greater extent as imagined, even when they have had a positive experience. However, respondents that have negative experiences will tend to tell others more than those respondents that have positive experiences. Overall, more respondents were likely to tell others about a negative experience than a positive experience, contrary to Naylor and Kleiser (2000). The role of the trust of business owners by respondents suggests that those that neither trust nor distrust business owners tend to engage in negative word of mouth, implying that their neutral trust is probably ambivalence. The large numbers of respondents that indicated that they do not tell others about their experience validates the observation by Yoshi and Musalem (2021) about underreporting, and suggests that people choose not to share this information for reasons that are uncertain. Positive word of mouth is between 17% and 18% based on the level of trust of business owners, whether neutral or definitely and somewhat trusting. Just over 12% of distrustful respondents will tell others about their negative experiences. These observations suggest that positive word of mouth has an ever slight advantage over negative word of mouth. Chi square analysis revealed that positive word of mouth has a statistically significant relationship to region of residence and level of education, partially validating the observations of Falls and Francis (1996), with no significant relationships among the variables with respect to negative word of mouth. Logistic regression analysis revealed that while none of the predictor variables were statistically significant for negative word of mouth behavior, the level of income, education level, and region of residence were statistically significant for positive word of mouth, partially validating the observations of Falls and Francis (1996) and Mladenović, Bruni, and Kalia (2021). These findings mainly imply that the likelihood to engage in positive word of mouth behavior varies across different regions of the United States, and increases with level of education, generally validating the broad observations of Singh and Sarma (2018)

With regard to the overall tendencies of specific demographic groups, the findings provide valuable insights. While respondents that were 55 years or older were more likely to tell others about positive and negative experiences, respondents 18 to 24 years old were more likely to tell others about negative experiences. Those respondents earning $30,000 or less were more likely to tell others about both positive and negative experiences. Females were equally likely to tell others about both positive and negative experiences. Positive word of mouth was more likely to happen with respondents from the Middle Atlantic region, and negative word of mouth was more likely with respondents from the South Atlantic region. Well-educated respondents with a master's degree tended to use positive word of mouth but those with a bachelor's degree or some college were more likely to use negative word of mouth

There are implications for business practice. First, word of mouth as an element of a firm's marketing strategy must be carefully considered. Given that the findings suggest that most people do not actively engage in word of mouth, businesses must learn how to activate interpersonal communication among consumers to create authentic positive promotion. Therefore, businesses must have a deep understanding of not just how their potential clients receive messaging but how to inspire them to share messaging on a personal level. For example, a "tell your friends" promotion that offers a discount or some valuable consideration for customers could change the word of mouth dynamic, and result in loyalty. Second, since the findings illustrate that there is a large group of consumers that don't explicitly trust business owners, this implies that not all consumers are willing to be loyal for some reason. Marketers must understand how to activate trust among these consumers to the extent that they would be willing to tell others about positive experiences. Therefore, marketers and businesses must understand the pain points of consumers that prevent them from ultimately being word of mouth promoters. Finally, the regional and demographic variations of word of mouth illustrate that a generic approach to inspiring word of mouth will not work with everyone nor everywhere. Greater consideration must be given to the consumer nuances in different markets and within market segments so that word of mouth messaging can be inspired, and addresses the right people.

## 5. Directions for future research

First, it is recommended that the findings of distrust or ambivalence toward business owners must be studied so that there is a deeper understanding of why consumers may be reluctant to be word of mouth promoters of businesses. Second, most research must be done with respect to the activation of word of mouth among consumers that choose not to share their experiences.

## 6. Limitations

Limitations must be considered in every research endeavor. First, although Amazon Mechanical Turk can allow researchers to collect data faster than ever before, there are problems associated with its use (2024b)  As a result, the author acknowledges that certain shortcomings have been considered in the use of mTurk (Berry, 2024b). Second, since respondents answered a multiple selection question wherein one of the choices was "tell others" in mu questionnaire (Berry, 2024a), the motivation for this choice was not explored in my dissertation. Finally, the author states that these findings are presented as synthesized knowledge from my dissertation, and therefore, are valuable insights for marketers and researchers.

## 7. Patents

There are no patents resulting from the work reported in this manuscript.

## 8. Funding

This research received no external funding.

## 9. Conflicts of Interest

The authors declare no conflict of interest.

## 10. Declaration of generative AI in scientific writing

The author declares that generative AI tools were not used in the writing or research of this article.